\documentclass[times, twoside]{StyleArXiv}
\usepackage{booktabs}
\usepackage{multicol,multirow}
\usepackage{orcidlink}
\usepackage{subcaption}
\usepackage{cleveref}
\crefname{figure}{Fig}{Figs}
\crefname{table}{Tab}{Tabs}
\crefname{equation}{eq.}{eqs.}

\leadauthor{Siddhad} 

\begin{document}

\title{Neural Networks Meet Neural Activity: Utilizing EEG for Mental Workload Estimation}
\shorttitle{Neural Networks Meet Neural Activity: Utilizing EEG for Mental Workload Estimation}

\author[1,\Letter]{Gourav~Siddhad~\orcidlink{0000-0001-5883-3863}}
\author[1]{Partha~Pratim~Roy~\orcidlink{0000-0002-5735-5254}}
\author[2]{Byung-Gyu~Kim~\orcidlink{0000-0001-6555-3464}}
\affil[1]{Department of Computer Science and Engineering, Indian Institute of Technology, Roorkee, Uttarakhand, 247667, India}
\affil[2]{Division of AI Engineering, Sookmyung Women's University, Seoul 04310, Republic of Korea}

\maketitle


\begin{abstract}
Electroencephalography (EEG) offers non-invasive, real-time mental workload assessment, which is crucial in high-stakes domains like aviation and medicine and for advancing brain-computer interface (BCI) technologies. This study introduces a customized ConvNeXt architecture, a powerful convolutional neural network, specifically adapted for EEG analysis. ConvNeXt addresses traditional EEG challenges like high dimensionality, noise, and variability, enhancing the precision of mental workload classification. Using the STEW dataset, the proposed ConvNeXt model is evaluated alongside SVM, EEGNet, and TSception on binary (No vs SIMKAP task) and ternary (SIMKAP multitask) class mental workload tasks. Results demonstrated that ConvNeXt significantly outperformed the other models, achieving accuracies of 95.76\% for binary and 95.11\% for multi-class classification. This demonstrates ConvNeXt's resilience and efficiency for EEG data analysis, establishing new standards for mental workload evaluation. These findings represent a considerable advancement in EEG-based mental workload estimation, laying the foundation for future improvements in cognitive state measurements. This has broad implications for safety, efficiency, and user experience across various scenarios. Integrating powerful neural networks such as ConvNeXt is a critical step forward in non-invasive cognitive monitoring.
\end{abstract}


\begin{corrauthor}
g\_siddhad\at cs.iitr.ac.in
\end{corrauthor}


\section{Introduction}
\label{sec_intro}

With its capacity to directly record the complex electrical activity of the brain, electroencephalography (EEG) is a potent technique for non-invasively assessing mental workload in real-time. Evaluating mental workload is crucial in high-stakes fields like medicine and aviation, but it's also important in regular workplaces and educational settings~\cite{raduntz2020indexing}. This non-invasive method is essential for improving brain-computer interface (BCI) applications and understanding neural processes. Its applications range from ensuring safety in critical operations to optimizing cognitive performance and tailoring learning experiences. Within the quickly developing field of cognitive neuroscience, EEG is essential for bridging the knowledge gap between theoretical concepts and real-world applications related to mental stress~\cite{das2024cognitive}. By examining the electrical patterns obtained from EEG, researchers can gain crucial knowledge about healthy and pathological brain functions~\cite{aznan2021leveraging}. This improves the capacity to control and interpret cognitive loads in various scenarios. This finding highlights EEG's dual potential to advance scientific understanding and practical application. These capabilities underline the EEG's critical role in the realms of current neuroscience and BCI research.

EEG, with its high temporal resolution, is a valuable tool for capturing rapid changes in mental workload. However, its high dimensionality, intrinsic noise, and non-stationarity~\cite{chakladar2021eeg} make it challenging to extract meaningful information. Additionally, the complexity of the human brain and limitations of EEG technology, such as variations in cognitive abilities, low signal-to-noise ratio, and poor spatial resolution~\cite{kostas2021bendr}, complicate the accurate localization of neural sources. Individual differences in brain responses and EEG acquisition methods further contribute to signal variability. To address these challenges and fully leverage EEG's potential in high-stakes situations, advanced machine learning models are crucial~\cite{rojas2021prediction}. These models can improve the accuracy and consistency of EEG data processing, facilitating the development of user-friendly systems for estimating mental workload. By incorporating subject-specific information, such as task complexity and individual cognitive characteristics, these models can enhance the classification accuracy of mental workload states~\cite{gupta2021subject}. This multidisciplinary approach is expected to drive significant advancements in the field, enabling more effective monitoring and analysis of cognitive processes.

Enter ConvNeXt~\cite{liu2022convnet}, a cutting-edge convolutional neural network (CNN) architecture initially designed to address computer vision problems. ConvNeXt is modified for EEG analysis in this work, which is a significant innovation. ConvNeXt has the potential to revolutionise EEG data processing by extracting subtle patterns from EEG signals more precisely and effectively than conventional models because of its improved convolutional operations, optimised layer structures, and effective training methodologies. This modification holds great potential for precisely detecting the complex patterns in brain signals that correlate to different mental workloads. It represents a shift from conventional EEG analysis techniques and could result in workload estimates that are more accurate and insightful. Because of ConvNeXt's improved feature extraction capabilities, neural signatures associated with different brain states, clinical diseases, and cognitive tasks may be precisely identified. This expands the field of neuroscience research and advances BCI technology, with broad implications ranging from better seizure detection to enhanced sleep stage categorization and cognitive load evaluation.

This study investigates the integration of the ConvNeXt architecture into EEG data analysis. It highlights the potential of ConvNeXt to revolutionize the precision of computational neuroscience and BCI research. Researchers can gain profound insights into the brain's electrical activity across various cognitive states by analysing EEG data. ConvNeXt's advanced design efficiently learns complex patterns within EEG data, outperforming traditional models in feature extraction. The use of ConvNeXt tackles the challenges associated with EEG signals, offering significant improvements in cognitive performance and ergonomics across diverse settings. Its potential to differentiate between types of cognitive loads sets a new standard for EEG analysis, marking a substantial advancement in the field. As research continues, the application of ConvNeXt in EEG analysis is expected to drive breakthroughs in understanding the brain's complexities and advancing cognitive ergonomics.

Contributions of this work are:
\begin{itemize}
    \item While CNNs have shown promise in EEG classification, ConvNeXt's advanced architecture ushers in a new era for efficiently extracting complex patterns from EEG data, potentially surpassing traditional models by a substantial margin.
    \item Beyond classification accuracy, techniques to interpret the learned features within the ConvNeXt model provide valuable insights into the specific neural activity patterns associated with different cognitive states, offering a deeper understanding of brain processes during controlled experiments.
    \item This work aims to significantly improve the current standards of EEG analysis, offering deeper insights into brain activity and paving the way for advancements in cognitive research.
\end{itemize}

The paper is structured as follows: Section~\ref{sec_related} reviews research on mental workload and EEG analysis, highlighting the challenges and opportunities. Section~\ref{sec_method} delves into the ConvNeXt model and its adaptation for EEG, including the description of the dataset and classifiers. Section~\ref{sec_experiment} presents the experimental setup and the details of the classifiers used. Section~\ref{sec_result} presents experimental findings, and Section~\ref{sec_conclusion} concludes this work and discusses future directions.


\section{Related Work}
\label{sec_related}

Recent research on estimating mental workload using EEG signals has produced various innovative methods and insights. Early studies, such as those by Hernandez et al.~\cite{hernandez2022recognition}, explored the evaluation of pilots' mental workloads in high-risk cockpit environments through multitasking. The work by Di et al.~\cite{di2018eeg} extended EEG analysis to the driving context, integrating EEG data with subjective assessments and vehicle dynamics to study the effects of traffic and road conditions on driver workload. Kartali et al.~\cite{kartali2019real} contributed by focusing on real-time mental workload estimation using EEG. Singh et al.~\cite{singh2023improved} applied a combination of 1D-CNN and Synthetic Minority Oversampling Technique (SMOTE) to enhance the classification accuracy of mental workload levels.

Advancements in machine learning have substantially improved the application and understanding of EEG technology in environments characterized by high stress and multitasking. Qu et al.~\cite{qu2022mental} addressed the challenge of EEG signal non-stationarity through cross-session subspace alignment, significantly improving signal classification across sessions. Mastropietro et al.~\cite{mastropietro2023reliability} highlighted the importance of electrode configurations and signal processing techniques in enhancing the sensitivity and reproducibility of EEG-based mental workload measurements. In their studies, So et al.~\cite{so2017evaluation} employed Support Vector Machines (SVM) to accurately estimate workload levels, demonstrating the efficiency of a single-channel EEG device in monitoring dynamic changes with notable accuracy. Similarly, Chin et al.~\cite{chin2018eeg} provided evidence that EEG can effectively differentiate between various levels of cognitive workload during mental arithmetic tasks, illustrating the method's versatility and effectiveness.

Furthermore, various machine learning algorithms, including SVM, Random Forest, and k-Nearest Neighbors (k-NN), have been applied to analyze EEG signals. Research by Singh et al.~\cite{singh2021mental} and Pandey et al.~\cite{pandey2020mental} has demonstrated varied success rates, particularly in multitasking scenarios where the analysis of statistical and fractal dimension (FD) features is crucial, as shown by Lim et al.~\cite{lim2015eeg}. These techniques have proven particularly effective in enhancing the practical applications of EEG in diverse cognitive studies.

Innovative computational approaches have greatly enhanced EEG-based classification tasks by integrating advanced deep-learning models~\cite{siddhad2024enhancing}. Cheng et al.~\cite{cheng20233d} introduced a novel combination of 3D-CNN with LSTM and attention mechanisms, significantly improving spatial-temporal feature learning. Similarly, Yao et al.~\cite{yao2024emotion} demonstrated the efficacy of combining CNN and transformer models, which has significantly advanced EEG-based classification tasks. Further contributions in this field include the work of Aldawsari et al.~\cite{aldawsari2023optimizing}, who optimized a 1D-CNN model, showcasing the potential of lightweight deep learning methods for real-time EEG-based emotion recognition. Saleh et al.~\cite{saleh2023efficient} further leveraged transformer networks to enhance classification accuracy on eye direction, PPG, and EEG data, demonstrating the versatility of transformers in handling various types of physiological data.

Additionally, Siddhad et al.~\cite{siddhad2024efficacy} utilized transformer networks for classifying raw EEG data, illustrating their adaptability to tasks beyond natural language processing, such as mental workload classification. This adaptation addresses common challenges in EEG data classification, including the dependency on pre-processing and the need for hand-crafted feature extraction, by leveraging deep learning to potentially automate these processes. Moreover, Parveen et al.~\cite{parveen2023attention} introduced an attention-based 1D-CNN for mental workload classification. This model enables the identification of specific patterns of brain activity associated with various workload levels, highlighting the precise capabilities of modern computational models in interpreting complex neural signals. These developments collectively push the boundaries of EEG data analysis, paving the way for more accurate and efficient applications in various fields.

In addition to machine learning, using artificial neural networks (ANN) in EEG analysis has been prominently featured, with research by Samima et al.~\cite{samima2019eeg} demonstrating notably high accuracy in estimating mental workload in operators. This achievement illustrates the increasing integration of complex neural network architectures in EEG data analysis, highlighting a significant trend in the field. This trend is further contextualized within the broader scope of advancements discussed in the `Brain Informatics' collection by Liu et al.~\cite{liu2023brain}. This collection covers a vast range of studies in brain science, human information processing systems, and brain big data analytics. It points to integrating advanced computational models like transformers in EEG analysis as part of a broader exploration of brain informatics technologies. Such integration is pivotal in advancing a comprehensive understanding of mental health through informatics paradigms, demonstrating the interconnection between innovative technological applications and fundamental brain science research.

Measurement, classification, and understanding of cognitive load have advanced significantly due to studies on EEG signals for mental workload estimation, especially in high-stress, multitasking situations. These researches have played a pivotal role in developing systems that evaluate cognitive load in real-time across various areas, including workplace safety, aviation, education, and the automotive industry. This has led to a deeper understanding of EEG and its practical applications. To fully utilize EEG-based workload estimation, several obstacles remain. These include the need for more in-depth task analysis that considers the complexities of real-world scenarios, advancements in real-time processing systems for increased accuracy and reduced latency, and the requirement for personalized models to address significant inter-subject variability and enhance accuracy.

Additionally, it is imperative to identify and optimize predictive EEG features using advanced machine-learning techniques. Integrating EEG data with other physiological indicators can provide a more thorough workload assessment. Models must also adapt to dynamic workload levels over prolonged tasks and consider individual cognitive differences and mental states for more precise estimations. Lastly, improving the usability and wearability of EEG devices is essential to facilitate their broader adoption, especially in workplace settings where practicality and comfort are critical. These collective efforts highlight the crucial role of advanced computational models and machine learning techniques in refining EEG applications and addressing their challenges to maximize their effectiveness across various fields.

Improving EEG-based mental workload estimation requires addressing the intrinsic challenges of EEG data, namely, high dimensionality, noise susceptibility, and non-stationary nature. Recurrent neural networks (RNNs) and convolutional neural networks (CNNs) are examples of advanced computational models necessary for precisely interpreting EEG data in precision-critical applications like brain-computer interfaces and medical diagnosis. These models effectively manage data from multiple scalp electrodes through dimensionality reduction and automatic feature extraction, increasing analytical accuracy and efficiency~\cite{ding2022tsception}. Moreover, EEG readings can be distorted by noise and artefacts from electrical interference, muscles, and other sources, which can mask brain activity. 
To overcome these problems, models like EEGNet use advanced noise-reduction techniques, which produce more reliable analyses~\cite{craik2019deep}. The analysis is further complicated by the non-stationary nature of EEG data, which can be impacted by shifts in the subject's physiological or cognitive state, task involvement levels, or environmental variables. Adaptive filtering or time-frequency analysis techniques are essential for capturing the dynamic elements of EEG signals because traditional signal processing approaches, which presume stationarity, are insufficient~\cite{vidaurre2011biosig}. Improvements in these areas can greatly increase the responsiveness, accuracy, and ease of use of EEG-based mental workload estimation, hence increasing its applications.

Incorporating advanced computational models is critical to effectively handle the intricacies of EEG data and improve applications in domains where accuracy is paramount, such as neurofeedback therapy. These models enable a robust and reliable method of EEG analysis by combining insights from computer science, cognitive science, and neuroscience. Essentially, using these models to read EEG signals reliably improves the interpretation of complex data, leading to more dependable and efficient applications in various fields.


\section{ConvNeXt}
\label{sec_method}

\begin{figure*}[!t]
	\centering
	\begin{subfigure}{0.17\linewidth}
		\includegraphics[width=\linewidth]{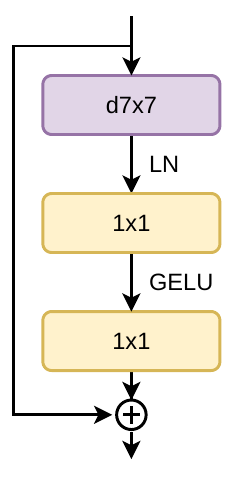}
		\caption{}
		\label{fig_method_convnextblock}
	\end{subfigure}
    \hfill
	\begin{subfigure}{0.78\linewidth}
		\includegraphics[width=\linewidth]{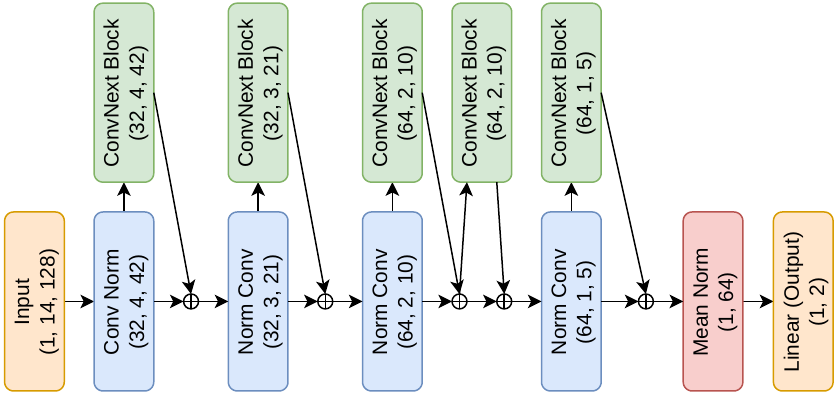}
		\caption{}
		\label{fig_method_convnext}
	\end{subfigure}
	\caption{(a) ConvNeXt Block and (b) ConvNeXt Model. Conv, Norm, and Mean in the ConvNeXt model represent 2D Convolution, Normalization, and Mean layer, respectively. The shape in each block represents the output shape of that block.}
	\label{fig_method}
\end{figure*}

This study incorporates a cutting-edge CNN, the ConvNeXt architecture~\cite{liu2022convnet}, into EEG-based measures of mental workload. ConvNeXt was originally developed for computer vision, but its ability to effectively learn complicated patterns within noisy and high-dimensional data makes it a promising tool for EEG analysis. The success of this hybrid architecture in image-related tasks can be attributed to its combination of the capabilities from both attention-based mechanisms of Transformer models and classic CNNs. The well-known ResNet design, which is renowned for its residual connections that allow the training of extremely deep networks, serves as an inspiration for ConvNeXt. ConvNeXt, however, differs greatly due to key modifications. It uses larger kernel sizes in its depthwise convolutions for wider receptive fields, layer normalisation for increased stability, and inverted residual structures to optimize computational efficiency. These modifications make ConvNeXt very well-suited to handle the complexities of EEG signals.

The ConvNeXt architecture is redesigned using convolutional blocks that gradually downsample input images while boosting channel capacity, structuring it into multiple stages suitable for EEG analysis. This modification improves the model's capacity to extract workload-related patterns by addressing the unique challenges of EEG data, such as high dimensionality, noise levels, and subtle signal fluctuations. Updated convolutional layers, stochastic depth, data augmentation, and LayerNorm, which stabilise the learning process, are some major modifications. Due to its scalability and processing efficiency, ConvNeXt is suited for real-time applications such as mental workload evaluation, sleep stage classification, and seizure detection. Through techniques like extensive data augmentation and a revised learning rate schedule to prevent overfitting and increase convergence, the combination of this advanced architecture with EEG data has the potential to revolutionise analysis, enhancing precision and extending the spectrum of applications.

The ConvNeXt architecture introduces several modifications over traditional CNNs to enhance its adaptability. Key to these adaptations is the replacement of all batch normalization with layer normalization. Unlike batch normalization, which standardizes inputs using batch mean and variance, layer normalization standardizes inputs across each feature for every data point, providing consistency regardless of batch size variations. In ConvNeXts, the AdamW optimizer is employed, which refines the standard Adam approach by applying accurate weight decay, thus improving regularization and generalization capabilities. It is further distinguished by its use of $1 \times 1$ conv layers, with the depth-wise conv layer repositioned at the top of the stack, unlike in ResNeXt blocks. Additionally, the Gaussian Error Linear Unit (GELU), a smoother variant of the ReLU, is utilized, enhancing the network's non-linear processing capabilities.

Significant architectural scaling was undertaken for ConvNeXt, as illustrated in Fig~\ref{fig_method_convnext}, for its application to EEG, where the original ConvNeXt dimensions were tailored for $224 \times 224$ image sizes. Specifically, the arrangement of ConvNeXt blocks was altered from $(3, 3, 9, 3)$ to $(1, 1, 2, 1)$, and the number of channels in convolutional layers was decreased from $(96, 192, 384, 768)$ to $(32, 32, 64, 64)$. These changes reduced the model's complexity and the number of parameters and decreased training time, enhancing efficiency. The adapted model was subsequently trained for 100 epochs, optimizing it for the specific challenges and requirements of EEG data analysis as applied in this context.

The success of this ConvNeXt adaptation to EEG has the potential for broader applications in biomedical engineering and cognitive neuroscience, where precise pattern recognition is crucial. This approach significantly advances deep learning applications in neurological assessment, bridging the gap between high-level computer vision techniques and EEG signal analysis.


\section{Experiments}
\label{sec_experiment}

\subsection{Experimental Data}
To validate the performance of the proposed methodology, it was tuned and applied to the open-access mental workload dataset known as the simultaneous task EEG workload (STEW) dataset~\cite{lim2018stew}. It consists of raw EEG data from 48 subjects who participated in a multitasking workload experiment that utilized the simultaneous capacity (SIMKAP) multitasking test. The signals were captured using the Emotiv EPOC EEG headset, with 16-bit A/D resolution, 128 Hz sampling frequency, and 14 channels, namely AF3, F7, F3, FC5, T7, P7, O1, O2, P8, T8, FC6, F4, F8, AF4 according to the 10-20 international system with two reference channels (CMS, DRL). There are two parts to the experiment. First, the data was acquired for 2.5 minutes with subjects at rest or ``No task''. Next, subjects performed the SIMKAP test with EEG being recorded and the final 2.5 minutes were used as the workload condition. Subjects rated their perceived mental workload on a rating scale of 1-9 after each experiment segment.

EEG signals in their raw form (captured from a device) contain noise and artefacts and must be cleaned before use. EEG data is imported and bandpass filtering is done to remove environmental/muscle noise from scalp EEG. After epoching and removing bad epochs from the data, independent component analysis (ICA) is applied and bad channels are manually removed. The dataset is used with a sampling rate of 128 Hz, the same as during acquisition. After min-max scaling, the data was epoched into one-second intervals with a half-second overlap, resulting in data shaped as (1, channel count, EEG length), i.e., (1, 14, 128), yielding a total of 26,910 samples. The dataset is split into 70:15:15 ratios for train, validation, and test sets.


\subsection{Experimental Setup}
The experimental setup involved a DELL Precision 7820 Tower Workstation, with Ubuntu 22.04 OS, Intel Core(TM) Xeon Silver 4216 CPU, and an NVIDIA RTX A2000 12GB GPU. This hardware facilitated the implementation of DL models using Python 3.10 and the PyTorch library. The Adam optimizer, known for its computational efficiency, was used with default parameters ($\eta$ = 0.001, $\beta_1$ = 0.9, $\beta_2$ = 0.999). EEGNet and TSception were trained for 100 epochs, with batches of 16 and a learning rate of $1e-4$. The Radial Basis Function (RBF) kernel from scikit-learn~\cite{sklearn} was used with default settings for SVM. Classification accuracy was determined through stratified five-fold cross-validation, averaging the results for comprehensive assessment.


\subsection{Classifiers}
This study uses four popular models for EEG analysis, namely EEGNet, TSception, Transformer, and SVM. EEGNet~\cite{lawhern2018eegnet} is a compact convolutional neural network specifically designed for EEG data. Its success in a range of EEG tasks, efficiency, and reduced complexity make it appropriate for smaller datasets, which has led to a rise in its use in both research and practical applications. A deep learning model designed for time-series EEG data, TSception~\cite{ding2022tsception} highlights the temporal dynamics in the data. It is skilled at capturing the intricacies found in EEG data because it can effectively extract temporal information at different scales. TSception has demonstrated efficacy in identifying emotions and evaluating cognitive burden. The Transformer~\cite{siddhad2024efficacy} is a neural network architecture that relies on self-attention mechanisms to compute representations of input sequences. By avoiding recurrence and convolution, the Transformer is highly parallelizable and efficient, capturing long-range dependencies in the data. Its encoder-decoder structure allows for sophisticated processing and understanding of complex EEG patterns. A popular machine learning algorithm used in EEG analysis is SVM~\cite{cortes1995support}. By determining the ideal distance in high-dimensional space between classes, SVMs are excellent at classification. Robustness is one of SVM's strongest points; it performs particularly well with high-dimensional EEG data and can handle non-linear correlations using kernels. Each method, TSception, SVM, and EEGNet, offers unique advantages for EEG analysis. Neural network-based, EEGNet and TSception are especially good at processing raw EEG data and automatically extracting useful features. At the same time, SVM offers a robust classification method with well-understood theoretical foundations.


\section{Performance Evaluation}
\label{sec_result}

\begin{table*}[!t]
    \centering
    \caption{Performance Comparison on STEW dataset with 95\% Confidence Interval}
    \label{tab_result}
    \begin{tabular}{l c c}
        \toprule
        \multirow{2}{*}{\textbf{Classifier}} & \textbf{No vs SIMKAP task} & \textbf{SIMKAP multi-task} \\
        & \textbf{2 class} & \textbf{3 class} \\
        \midrule
        \textbf{SVM}~\cite{cortes1995support} & $83.34 \pm 0.39$ & $83.21 \pm 0.28$ \\
        \textbf{EEGNet}~\cite{lawhern2018eegnet} & $84.33 \pm 1.03$ & $83.64 \pm 1.33$ \\
        \textbf{TSception}~\cite{ding2022tsception} & $95.21 \pm 0.53$ & $94.73 \pm 0.30$ \\
        \textbf{Transformer}~\cite{siddhad2024efficacy} & $95.32 \pm 0.00$ & $88.72 \pm 0.00$ \\
        \midrule
        \textbf{Proposed ConvNeXt} & $\mathbf{95.76 \pm 0.51}$ & $\mathbf{95.11 \pm 0.80}$ \\
        \bottomrule
    \end{tabular}
\end{table*}

\begin{figure*}[!t]
	\centering
	\begin{subfigure}{0.48\linewidth}
		\includegraphics[width=\linewidth]{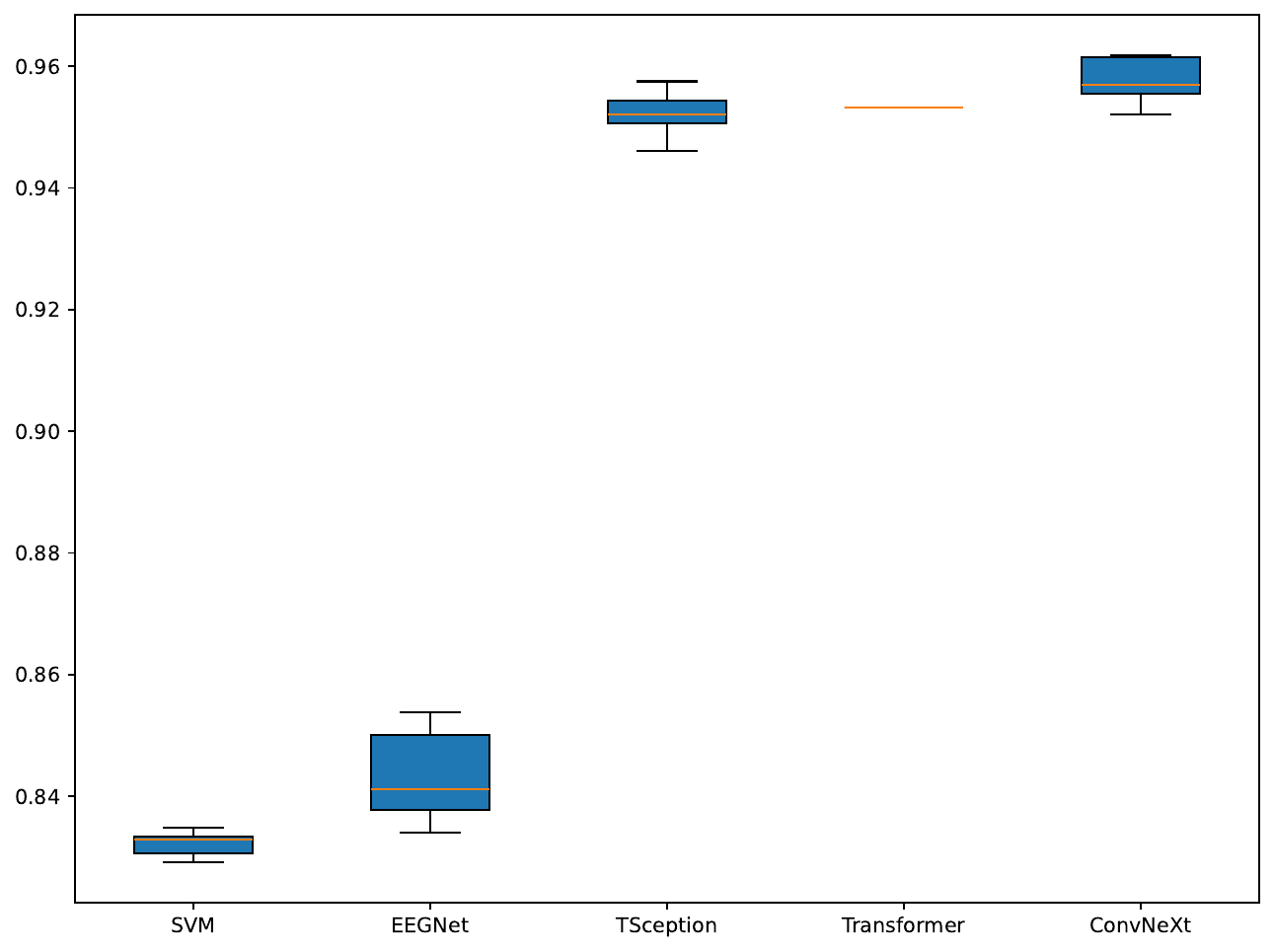}
		\caption{}
		\label{fig_boxplo3}
	\end{subfigure}
    \hfill
	\begin{subfigure}{0.48\linewidth}
		\includegraphics[width=\linewidth]{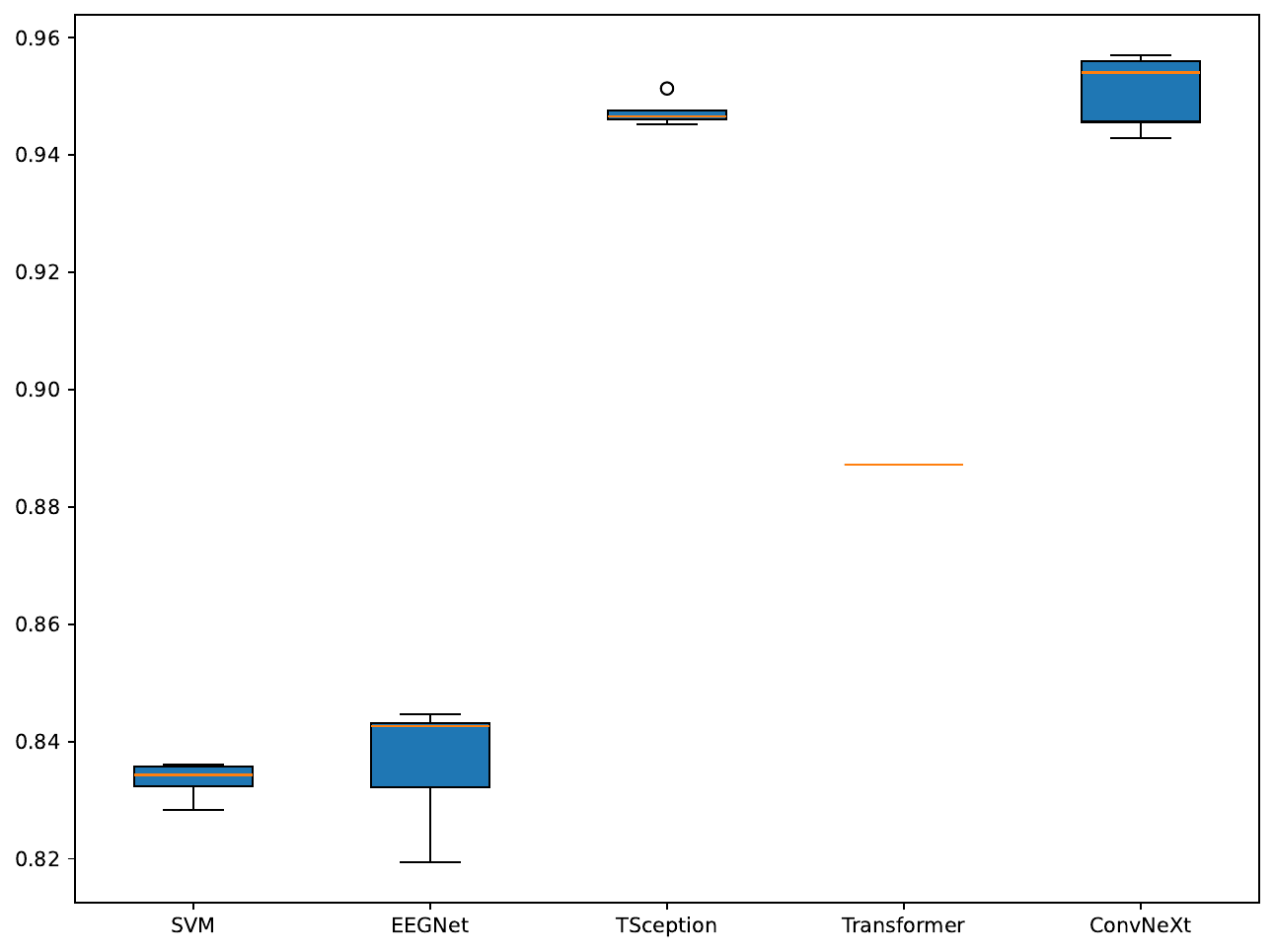}
		\caption{}
		\label{fig_boxplot2}
	\end{subfigure}
	\caption{Box plot for (a) `No vs SIMKAP task' and (b) `SIMKAP multi-task'.}
	\label{fig_boxplot}
\end{figure*}

This research aims to improve understanding of cognitive states by estimating mental workload through EEG, an essential tool in cognitive neuroscience. The effectiveness of several advanced machine learning classifiers was evaluated on the STEW dataset, a benchmark in cognitive load research. The study involved two primary analyses: a binary classification task (No vs. SIMKAP) and a more complex multi-class task (SIMKAP multi-task), designed to test the classifiers under varying complexities of mental workload representation. Five classifiers were assessed: SVM, EEGNet, TSception, Transformer, and a novel classifier introduced in this study. Their performance was measured in terms of accuracy and reliability. Results were substantiated by 95\% confidence intervals to confirm the robustness of the findings.

The results demonstrate significant contributions to the discourse on neural network-based approaches and machine learning techniques in EEG analysis. The ConvNeXt model, in particular, showcased exemplary performance in accurately classifying mental workload levels under diverse conditions. This achievement sets new standards for accuracy and reliability in mental workload estimation. The analysis methodically presents classification accuracies, highlighting the technical strengths of this approach and its broader implications for cognitive neuroscience and human-computer interaction.

The evaluation methodology utilizes the STEW dataset, organized into two scenarios, to assess the model's classification capabilities thoroughly. In the binary classification scenario, the model distinguishes between a no-task baseline and a defined cognitive task, known as the 'No vs SIMKAP task.' This scenario is critical for evaluating the model's accuracy in differentiating resting cognitive states from those involved in a SIMKAP task. The more complex ternary classification scenario introduces a gradient of mental workload levels—classified as low, medium, and high, together referred to as 'SIMKAP multitask.' This detailed evaluation demonstrates the model's proficiency in differentiating among varied cognitive loads, mirroring real-world conditions as depicted by the multidimensional data of the STEW dataset.

Table \ref{tab_result} presents a comparative analysis of five classifiers: SVM, EEGNet, TSception, Transformer, and the proposed ConvNeXt model, across the two classification scenarios. The binary task involves two classes, while the multi-class task includes three. This table not only highlights the performance metrics of each classifier, presumably in accuracy percentages with 95\% confidence interval, underscoring variability in performance across different trials or datasets. This variability is crucial for evaluating the robustness and reliability of the models in real-world applications. Moreover, Table \ref{tab_result} emphasizes the effectiveness of advanced machine learning and neural network models in interpreting EEG data for mental workload classification. Notably, TSception and the proposed method exhibit exceptional promise, as indicated by their strong performance metrics.

The results demonstrate the variability in performance metrics of different classifiers evaluated on the STEW dataset. The SVM classifier shows moderate effectiveness, achieving 83.34\% (±0.39) accuracy in the binary classification task and 83.21\% (±0.28) in the multi-class task. EEGNet exhibits slightly better performance, with 84.33\% (±1.03) accuracy for the binary task and 83.64\% (±1.33) for the multi-class task. TSception, a more advanced classifier, significantly outperforms the other models with impressive accuracies of 95.21\% (±0.53) in the binary classification and 94.73\% (±0.30) in the multi-class scenario. The Transformer model achieves high performance in the binary task with an accuracy of 95.32\% (±0.00) but shows a drop in performance for the multi-class task with an accuracy of 88.72\% (±0.00). The proposed model records the highest accuracies of 95.76\% (±0.51) in the binary task and 95.11\% (±0.80) in the multi-class task, demonstrating superior capability in managing the classification challenges of the STEW dataset.

These results underscore the comparative strengths and potential real-world applicability of advanced neural network-based classifiers, particularly TSception and the proposed model, in complex EEG data classification tasks. A detailed boxplot presented in Figure \ref{fig_boxplot} illustrates the distribution of classifier accuracies across multiple trials, providing a visual performance comparison. While TSception closely rivals the proposed model's effectiveness, it still performs exceptionally well, especially in EEG-based workload classification. Although outperformed by the more advanced classifiers, SVM maintains strong and reliable performance across both tasks. Similarly, EEGNet, despite being the least accurate of the tested classifiers, still holds its ground, particularly in the multi-class scenario. This visual evidence, as depicted in the boxplot, highlights the importance of choosing an appropriate classifier based on the specific requirements and complexities of the task, demonstrating that newer or more sophisticated models like TSception or the proposed model can offer substantial benefits in certain scenarios. The accuracy of the Transformer model is derived from the respective research paper that utilized the same dataset. However, it did not include the confidence interval, showing it as ±0.00.

Overall, the proposed model achieves the highest accuracy and robustness across both classification tasks, affirming its superiority in addressing the challenges presented by the `No vs SIMKAP' and `SIMKAP multi-task' scenarios. These findings represent a pivotal advancement in non-invasive methods for quantifying mental workload, providing insights that could improve real-time cognitive state assessment and enhance safety, efficiency, and user experience in high-demand settings. The efficacy of the ConvNeXt model, tailored for EEG signal analysis, underscores the potential of advanced machine learning classifiers in this field. Moving forward, this study sets the stage for further innovations in mental workload assessment using EEG data, aiming to refine and expand the applicability of these techniques in both theoretical and practical contexts.

By elucidating the strengths and limitations of current methodologies, this research also suggests directions for future efforts to improve and innovate in the area of mental workload estimation. The comparative analysis across different models aids in understanding the potential of neural network-based approaches for real-world applications, inspiring continued exploration and development in cognitive neuroscience and related fields.


\section{Conclusion}
\label{sec_conclusion}

This study assessed the classification capabilities of advanced neural network-based classifiers using the STEW dataset for binary and multi-class scenarios, aiming to evaluate each model's precision in distinguishing between a no-task baseline and SIMKAP tasks for neuroscientific applications. The proposed model displayed exceptional accuracy, demonstrating its potential for real-world neuroscientific settings, such as real-time mental workload assessments in critical sectors like air traffic management and healthcare. This could significantly enhance safety and efficiency.

The findings mark a significant advancement in mental workload estimation through EEG analysis, achieved using a customized ConvNeXt model. This model accomplished classification accuracies of 95.76\% for the binary (No vs SIMKAP task) and 95.11\% for the ternary (SIMKAP multitask) classes, showcasing the potential of deep learning in the sophisticated interpretation of EEG data. These results contribute substantially to fields like cognitive neuroscience and human-computer interaction.

Looking ahead, this study encourages further research to address current limitations by expanding dataset diversity and exploring broader practical applications. By reinforcing the importance of selecting appropriate classifiers and advancing machine learning in neuroscientific research, this work sets the stage for future studies to create more intuitive and adaptive interfaces, meeting the complex needs of diverse sectors. Further investigations should validate and potentially broaden the applicability of these models in practical scenarios.




\section*{Bibliography}
\bibliography{references}








\end{document}